\documentclass[11pt,a4paper]{article}
\pdfoutput=1
\usepackage[pdftex]{graphics}
\usepackage{jheppub}
\usepackage{amsmath,amssymb,amsfonts}
\usepackage{enumitem}
\usepackage{multirow}
\usepackage{array,booktabs}
\usepackage{slashed}
\usepackage[utf8]{inputenc}
\usepackage[dvipsnames]{xcolor}

\title{Poincar\'e generators at second post-Minkowskian order}

\author[a]{Hojin Lee}
\author[e,f]{Kanghoon Lee}
\author[a,b,c,d]{Sangmin Lee}

\affiliation[a]{Department of Physics and Astronomy, Seoul National University, Seoul 08826, Korea}
\affiliation[b]{Center for Theoretical Physics, Seoul National University, Seoul 08826, Korea}
\affiliation[c]{College of Liberal Studies, Seoul National University, Seoul 08826, Korea}
\affiliation[d]{School of Physics, Korea Institute for Advanced Study, Seoul 02455, Korea}
\affiliation[e]{Asia Pacific Center for Theoretical Physics, Pohang 37673, Korea}
\affiliation[f]{Department of Physics, Postech, Pohang 37673, Korea}

\emailAdd{zet4gra9er@snu.ac.kr, kanghoon.lee1@gmail.com, sangmin@snu.ac.kr}

\abstract{We verify the global Poincar\'e invariance of the Hamiltonian mechanics of gravitating binary dynamics at the second post Minkowskian (2PM) order. For spinless point particles, based on the known 2PM Hamiltonian in the center of momentum frame, we compute the general 2PM Hamiltonian valid in an arbitrary reference frame. 
An off-shell extension of the 1PM Hamiltonian, which contributes at the 2PM order through an iteration process, plays a crucial role.
We then construct the 2PM boost generator that uniquely satisfies all the conditions imposed by the Poincar\'e algebra. 
}

\begin{document}

\maketitle

\section{Introduction}

Ever since the pioneering work \cite{Damour:2000kk} by Damour, Jaranowski and Sch\"afer, 
global Poincar\'e generators have provided stringent consistency checks for post-Newtonian (PN) 
approach to the effective Hamiltonian mechanics of a gravitating binary system. 
At each order of the perturbative Hamiltonian, constructing the Poincar\'e generators (the boost generator in particular) 
can reconfirm the validity of the Hamiltonian and may even fix undetermined coefficients or detect errors.

In the context of post-Minkowskian (PM) expansion strongly influenced by scattering amplitudes (see {\it e.g.} \cite{Bjerrum-Bohr:2022blt,Kosower:2022yvp,Buonanno:2022pgc} for comprehensive reviews), the construction of the Poincar\'e generators 
was initiated only recently in \cite{Lee:2023nkx}, where the Hamiltonian $H^{[1]}$ and the boost generator $\vec{G}^{[1]}$, including all spin multipole moments, were constructed at the first post-Minkowskian order (1PM). 
The present paper takes a step further in the same direction.

To pinpoint the novelties of the 2PM computation, we first review the 1PM results \cite{Lee:2023nkx}. 
At 0PM (a pair of free particles), it is well known that 
\begin{align}
    H^{[0]} = E_1 + E_2 \,,
    \quad 
    \vec{G}^{[0]} = E_1 \vec{x}_1 + E_2 \vec{x}_2 \,, 
    \quad 
    E_a = \sqrt{\vec{p}_a^2 + m_a^2} \quad 
    (a=1,2) \,.
    \label{G-0PM}
\end{align}
The 1PM Hamiltonian is 
\begin{align}
\begin{split}
      H^{[1]} &=  \gamma_c \left[  - \frac{Gm_1^2 m_2^2(2\gamma^2 -1)}{E_1 E_2 r}  \right]\,, 
      \quad 
      \gamma = - \frac{p_1 \cdot p_2}{m_1 m_2} = \frac{E_1E_2 - \vec{p}_1 \cdot \vec{p}_2}{m_1m_2} \,, 
      \\
      \gamma_c &= \left[ 1-\vec{u}_c^2 + (\hat{n}\cdot\vec{u}_c)^2 \right]^{-1/2} \,,
      \quad 
      \vec{u}_c = \frac{\vec{p}_1 + \vec{p}_2}{E_1+E_2} \,, 
      \quad 
      \vec{r} = \vec{x}_1 - \vec{x}_2 \,,
      \quad 
      \hat{n} = \frac{\vec{r}}{r} \,.
\end{split}
\end{align}
The 1PM boost generator is 
\begin{align}
\begin{split}
    \vec{G}^{[1]} &= H^{[1]} \vec{X}^{[1]} \,,\quad \vec{X}^{[1]} = z_2 \vec{x}_1 + z_1 \vec{x}_2  \,,
    \quad 
    \quad 
    z_a = \frac{E_a}{E_1+E_2} \,.
\end{split}
 \label{G-1PM}
\end{align}
Since the boost generator takes one inertial frame to another, it is necessary to generalize the Hamiltonian to a form valid in 
an arbitrary ``lab" frame. At 1PM, the change of frame results in the ``dressing" factor $\gamma_c$ which is a measure for the deviation from the COM frame. The fact that $\vec{G}^{[1]}$ has a simple factorized form is another main finding of \cite{Lee:2023nkx}.

The 2PM Hamiltonian was first derived in the center of momentum (COM) frame in \cite{Cheung:2018wkq}; see also \cite{Damour:2017zjx,Bjerrum-Bohr:2018xdl,KoemansCollado:2019ggb,Cristofoli:2019neg,Cristofoli:2020uzm,Parra-Martinez:2020dzs,Bjerrum-Bohr:2021vuf}. 
It consists of four terms, 
\begin{align}
\begin{split}
      H^{[2]}|_\mathrm{COM} &= H^{[2]}_{1,\mathrm{b}} + H^{[2]}_{2,\mathrm{b}} + H^{[2]}_{3,\mathrm{b}} + H^{[2]}_{4,\mathrm{b}} \,, 
\end{split}
\end{align}
where the subscript b stands for ``bare".
From a diagrammatic point of view, $H^{[2]}_1$ 
originates from one-loop triangle diagrams. The other three are not directly produced by one-loop diagrams. 
Rather, they are remainders from an iteration of the 1PM Hamiltonian which cancels out the so-called super-classical term 
from the one-loop box diagrams.

One of the two main results of this paper is the general form of the Hamiltonian.
Going to the lab frame amounts to dressing the four terms in a few distinct ways, 
\begin{align}
\begin{split}
       &H^{[2]} = H^{[2]}_{1} +  H^{[2]}_{2} +  H^{[2]}_{3} + H^{[2]}_{4} \,, \quad \gamma_o = (1-\vec{u}_c^2)^{-1/2}  \,,
\\
   &H^{[2]}_{1} = \gamma_c^2 \gamma_o^{-1} H^{[2]}_{1,\mathrm{b}} \quad 
    H^{[2]}_{(2,3)} = \gamma_c^2 H^{[2]}_{(2,3),\mathrm{b}} \,,
    \quad 
     H^{[2]}_{4} = (3\gamma_c^2 - 2 \gamma_c^4) H^{[2]}_{4,\mathrm{b}} \,.  
\end{split}
\label{2pm-hamiltonian-dressing}
\end{align}
Understanding these dressing factors takes a large part of this paper. Here, we describe briefly the origin of the last term in \eqref{2pm-hamiltonian-dressing} which has no 1PM counterpart. 

As emphasized in {\it e.g.} \cite{Jones:2022aji}, scattering amplitudes are gauge invariant, on-shell quantities 
while potentials are gauge dependent, off-shell quantities. 
When deducing the form of a potential from an amplitude, the off-shell extension of the potential should be carefully spelled out, as discussed in the COM frame in the original work \cite{Cheung:2018wkq}. As we will explain in the main body of this paper, the iteration process in the lab frame produces an extra contribution to $H^{[2]}_4$ that is not visible in the COM frame.

Once the general form of the Hamiltonian is given, 
the search for the boost generator is a matter of straightforward (but lengthy) computation. 
There are three types of constraints, which we call $H$/$P$/$J$-conditions. 
Following \cite{Lee:2023nkx}, we propose an ansatz satisfying the $H$ condition 
and involving a small number of unknown functions, and then solve the $P$/$J$-conditions to determine the functions. 
The result is slightly more complicated than what one may naively expect from the 0PM \eqref{G-0PM} and the 1PM \eqref{G-1PM} expressions:
\begin{align}
    \vec{G}^{[2]} = H^{[2]} \vec{X}^{[1]} +  \gamma_c^2 H^{[2]}_{4,\mathrm{b}} z_{12} \vec{r} \,, 
    \quad 
    z_{12} = z_1 - z_2 \,.
    \label{G-2PM}
\end{align}
The reappearance of $\vec{X}^{[1]}$ and the ``misalignment" of 
$H^{[2]}_4$ with respect to the other three terms are the two most notable features of this result.

The main body of this paper consists of two sections. In section~\ref{sec:2pm-hamiltonian}, we explain how to derive the general 2PM Hamiltonian \eqref{2pm-hamiltonian-dressing}. In section~\ref{sec:2pm-boost}, we explain how to construct the boost generator \eqref{G-2PM}. 
We conclude in section~\ref{sec:discussion} with a few possible future directions. 
The two appendices provide some technical details for section~\ref{sec:2pm-hamiltonian} and \ref{sec:2pm-boost}.

\section{Hamiltonian} \label{sec:2pm-hamiltonian}

The 2PM COM Hamiltonian was first constructed in \cite{Cheung:2018wkq}; see also \cite{Damour:2017zjx,Bjerrum-Bohr:2018xdl,KoemansCollado:2019ggb,Cristofoli:2019neg,Cristofoli:2020uzm,Parra-Martinez:2020dzs,Bjerrum-Bohr:2021vuf}. 
The goal of this section is to find the general form of the 2PM Hamiltonian valid in an arbitrary lab frame. 

\paragraph{1PM} The 2PM Hamiltonian is intimately related to the 1PM Hamiltonian. 
To see the connection clearly, and to establish our notations, we quickly review the 1PM Hamiltonian. 
The 0PM Hamiltonian is the Minkowskian kinetic term, 
\begin{align}
    H^{[0]} = E_1 + E_2\,, \quad E_a = \sqrt{\vec{p}_a^2 + m_a^2} \,. 
\end{align}
The 1PM Hamiltonian in a lab frame is \cite{Lee:2023nkx}
\begin{align}
\begin{split}
      H^{[1]} &=  \gamma_c \left[  - \frac{Gm_1^2 m_2^2(2\gamma^2 -1)}{E_1 E_2 r}  \right]\,, 
      \quad 
      \gamma = - \frac{p_1 \cdot p_2}{m_1 m_2} = \frac{E_1E_2 - \vec{p}_1 \cdot \vec{p}_2}{m_1m_2} \,, 
      \\
      \gamma_c &= \left[ 1-\vec{u}_c^2 + (\hat{n}\cdot\vec{u}_c)^2 \right]^{-1/2} \,,
      \quad 
      \vec{u}_c = \frac{\vec{p}_1 + \vec{p}_2}{E_1+E_2} \,, 
      \quad 
      \vec{r} = \vec{x}_1 - \vec{x}_2 \,,
      \quad 
      \hat{n} = \frac{\vec{r}}{r} \,.
\end{split}
\end{align}
The transverse Lorentz factor $\gamma_c$ originates from the Fourier integral, 
\begin{align}
    4\pi \int_{\vec{q}} \,  \, \frac{e^{i\vec{q}\cdot\vec{r}}}{\vec{q}^2- (q^0)^2} = 
     4\pi \int_{\vec{q}}  \,  \, \frac{e^{i\vec{q}\cdot\vec{r}}}{\vec{q}^2- (\vec{u}_c\cdot \vec{q})^2 } = \frac{\gamma_c}{r} \,.
     \label{1PM-TL}
\end{align}
For later convenience, we introduce a few more shorthand notations, 
\begin{align}
   \vec{u}_a \equiv \frac{\vec{p}_a}{E_a} \,,
   \quad 
   z_a \equiv \frac{E_a}{E_1+E_2} \,,
   \quad 
    \vec{u}_- \equiv \vec{u}_1 - \vec{u}_2 \,,
    \quad 
    z_{12} \equiv z_1 - z_2  \,, 
    \quad 
    \xi \equiv z_1 z_2 \,.
\end{align}

\paragraph{2PM} \label{sec:ns-2pm-H}

The 2PM Hamiltonian in the COM frame \cite{Cheung:2018wkq} is given by 
\begin{align}
\begin{split}
      H^{[2]}_\mathrm{b} &= H^{[2]}_{1,\mathrm{b}} + H^{[2]}_{2,\mathrm{b}} + H^{[2]}_{3,\mathrm{b}} + H^{[2]}_{4,\mathrm{b}} = -\frac{G^2M m_1 m_2}{4 r^2} \left[ F_1 + F_2 + F_3 + F_4 \right] \,,
    \\
    F_1 &= 3 \left( \frac{m_1m_2}{E_1E_2}\right) (5\gamma^2 -1 ) \,,
    \qquad
    F_2 = -16 \left( \frac{m_1m_2}{E_1E_2}\right)^2 \frac{E}{M} \gamma(2\gamma^2-1) \,,
    \\
    F_3 &= 2 \left( \frac{m_1m_2}{E_1E_2}\right)^3 \frac{E}{M}  (2\gamma^2-1)^2 \,,
    \quad
    F_4 = -2 \frac{(m_1m_2)^3}{(E_1 E_2)^2} \frac{1}{E M} (2\gamma^2-1)^2 \,, 
    \\
    &\qquad M = m_1 + m_2 \,,
    \quad 
    E = E_1 + E_2 \,.
\end{split}
\label{H-2PM-COM} 
\end{align}
We pulled out an overall factor so that $F_{1,2,3,4}$ are dimensionless. The subscript b stands for ``bare". As we move to a lab frame, each term in the first line of \eqref{H-2PM-COM} will be ``dressed" by a factor that depends on $\vec{u}_c$ and $\hat{n}$. Before we discuss the dressing, we digress to give a quick review of how the four terms are determined by one-loop scattering amplitudes.

\begin{figure}[ht]
    \centering
    \includegraphics[width=11cm]{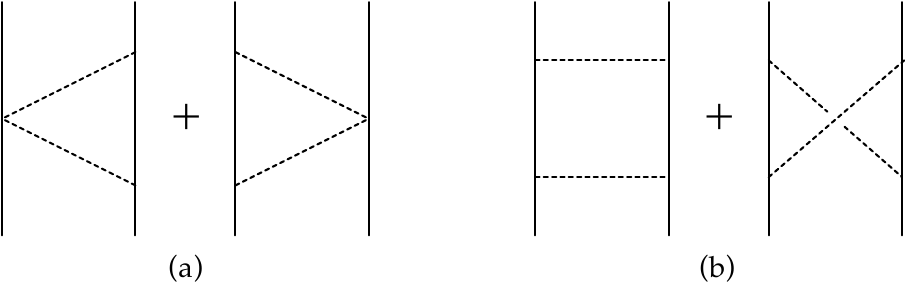}
    \caption{A simplistic depiction of one-loop diagrams relevant for the 2PM Hamiltonian. The solid external lines represent the massive bodies, and the dashed internal lines represent graviton exchanges. Only triangle and box diagrams contribute to the classical Hamiltonian; other diagrams only affect quantum corrections. Triangle diagrams determine $H^{[2]}_1$. In $D=4$, the box diagrams do not directly contribute to $H^{[2]}$, but the iteration of the 1PM Hamiltonian produces $H^{[2]}_{2,3,4}$.}
    \label{fig:1-loop-diagrams}
\end{figure}

The diagrammatic origin of the four terms in \eqref{H-2PM-COM} 
is depicted schematically in Figure~\ref{fig:1-loop-diagrams}. 
As far as the classical Hamiltonian is concerned, 
the only relevant diagrams are the triangle and box diagrams. 
Bubble and other diagrams may only affect quantum corrections. 
The sum of the two triangle diagrams directly determines $H^{[2]}_1$. The other three terms, $H^{[2]}_{2,3,4}$, are produced through an indirect route. 

In four space-time dimensions, the sum of the box and crossed box diagrams does not directly contribute to $H^{[2]}$. Instead, 
it produces an infrared divergent ``super-classical" term, 
which should be cancelled by an iteration of the 1PM Hamiltonian: 
a process commonly called ``Born subtraction". In the notation of \cite{Cristofoli:2019neg}, the subtraction process reads
\begin{align}
    \left\langle p^{\prime}|V| p \right\rangle=\mathcal{M}(p^{\prime}, p )-\int_{\vec{k}} \frac{\mathcal{M}(p^{\prime}, k) \mathcal{M}\left(k, p\right)}{E_p-E_k+i \epsilon}+\cdots \,,
    \label{born-subtraction}
\end{align}
where $\mathcal{M}$ is the on-shell amplitude from the quantum field theory and $V$ is the interacting part of the Hamiltonian. 
After cancelling the unphysical divergence, the Born subtraction produces $H^{[2]}_{2,3,4}$. To emphasize this indirect origin, 
and to prepare for later computations, 
we record the relation between $H^{[2]}_{2,3,4}$ and $(H^{[1]})^2$ before the dressing:
\begin{align}
\begin{split}
    H^{[2]}_{2,\mathrm{b}} & = \frac{E}{m_1 m_2} \frac{4\gamma}{2\gamma^2-1} ( H^{[1]}_\mathrm{b})^2 \,,
    \quad 
     H^{[2]}_{3,\mathrm{b}}  = -\frac{E}{2E_1 E_2} ( H^{[1]}_\mathrm{b})^2 \,,
    \quad 
    H^{[2]}_{4,\mathrm{b}}  = \frac{1}{2E} ( H^{[1]}_\mathrm{b})^2 \,.
\end{split}
\label{1PM-square-bare}
\end{align}

\paragraph{Dressing factor} 
The dressing factor of $H^{[2]}_{1}$ is easy to determine. 
The triangle diagrams are proportional to $|q|^{-1}$. 
In the lab frame, as noted in the 1PM setting \cite{Lee:2023nkx}, $q$ is the 4-vector satisfying $q\cdot(p_1+p_2)=0$ 
or $q^0 = \vec{u}_c\cdot \vec{q}$. 
The Fourier integral of $|q|^{-1}$ is then given by
\footnote{See appendix~\ref{sec:more-hamiltonian} for a collection of all Fourier transforms needed in this paper.}
\begin{align} \label{Fourier_2PM}
    2 \pi^2 \int_{\vec{q}} \frac{e^{i \vec{q} \cdot \vec{r}}}{\sqrt{\vec{q}^2- (\vec{u}_c\cdot \vec{q} )^2}} =\frac{ \gamma_c^2 \gamma_o^{-1} }{r^2} \,,
    \quad 
    \gamma_o = (1-\vec{u}_c^2)^{-1/2} \,.
\end{align}
It determines the dressing factor for $H^{[2]}_1$: 
\begin{align}
    H^{[2]}_1 = \gamma_c^2\gamma_o^{-1} \, H^{[2]}_\mathrm{1, b}\,.
    \label{2PM-dressing-1}
\end{align}

The iteration structure of $H^{[2]}_{2,3,4}$ 
translates to the fact that the $r^{-2}$ potential may come from a convolution integral. In the COM frame, the relevant integral is 
\begin{align}
    \frac{1}{r^2} 
    = (4\pi)^2 \int_{\vec{q}}  \int_{\vec{l}}\frac{e^{i \vec{q} \cdot \vec{r}}}{\vec{l}^2(\vec{q}-\vec{l})^2} \,.
\end{align}
Each factor in the denominator is a copy of the graviton propagator part of the tree amplitude. 
To generalize the integral to an arbitrary lab frame, it would be natural to modify the denominators as in \eqref{1PM-TL} and obtain
\begin{align}
    \frac{\gamma_c^2}{r^2} =  (4\pi)^2 \int_{\vec{q}}  \int_{\vec{l}}\frac{e^{i \vec{q} \cdot \vec{r}}}{(\vec{l}^2)_c (\vec{q}-\vec{l})^2_c} \,, \quad (\vec{a}\cdot\vec{b})_c \equiv \vec{a} \cdot\vec{b} - (\vec{u}_c \cdot \vec{a})(\vec{u}_c \cdot \vec{b}) \,.
    \label{denominator-naive}
\end{align}
If there were no extra contributions, we would be led to the conclusion that 
\begin{align}
     H^{[2]}_{(2,3,4)} = \gamma_c^2 \, H^{[2]}_\mathrm{(2,3,4) b} \quad (?) \,.
     \label{2PM-dressing-234}
\end{align}
As it turns out, the argument above gives the correct answer for $H^{[2]}_{2,3}$ but not for $H^{[2]}_{4}$. 
To recover the missing piece, we have to take a closer look at the Born subtraction.

\paragraph{Off-shell extension} 

Let $(\vec{p}_1, \vec{p}_2)$ be the incoming momenta and $(\vec{p}'_1 , \vec{p}'_2)$ be the outgoing momenta. 
The Born subtraction for the 2PM potential introduces an intermediate \emph{off-shell} state with momenta $(\vec{k}_1, \vec{k}_2)$.
The off-shell state strictly obeys the momentum conservation, 
\begin{align}
    \vec{p}_1 + \vec{p}_2  = \vec{P} = \vec{k}_1 + \vec{k}_2 \,.
    \label{off-shell-P}
\end{align}
But, by definition, it violates energy conservation, 
\begin{align}
    E_1(p) + E_2(p) = E_p \neq E_k = E_1(k) + E_2(k) \,.
    \label{off-shell-E}
\end{align}

A major intermediate step of the Born subtraction \eqref{born-subtraction} involves the integral,
\begin{align}
    \int_{\vec{k}} \frac{c_1(p',k) c_1(k,p)}{\left(E_p-E_k+i \epsilon\right)(\vec{p}'-\vec{k})^2_c (\vec{k}-\vec{p})^2_c} \,, 
    \quad c_1 \propto \frac{2\gamma^2-1}{E_1 E_2} \,.
    \label{Born-lab}
\end{align}
We recognize two copies of the 1PM potential as well as the ``propagator" $1/(E_p - E_k)$ from the non-relativistic quantum mechanics to be matched with the full quantum field theory 
in the classical limit. 

As explained in \cite{Cheung:2018wkq,Cristofoli:2019neg} for the COM Hamiltonian, 
for each numerator factor, we need to prescribe how to interpolate between the two momenta. For example, using the fact that $E_1$, $E_2$ and $\gamma$ are all functions of $\vec{p}^2 = \vec{p}_1^2 = \vec{p}_2^2$ only in the COM frame, we may assign 
\begin{align}
    c_1(k,p)_\mathrm{COM} =  c_1\left( \frac{k^2 +p^2}{2} \right) = c_1(p^2) + \frac{k^2-p^2}{2} c_1'(p^2) + \cdots \,. 
\end{align}
As we move to a lab frame, some complications arise. Most notably, the level surface of $E = E_1 + E_2$ for a fixed $\vec{P} = \vec{p}_1 + \vec{p}_2$ is no longer a sphere but an ellipsoid. Nevertheless, the process of extracting contributions 
from $c_1$ to $H^{[2]}_{2,3,4}$ remains largely unchanged aside from the dressing by $\gamma_c^2$ as in \eqref{2PM-dressing-234}. 

The extra contribution to $H^{[2]}_{4}$ comes from the denominator factors:
\begin{align}
    D(k,p) = \frac{1}{\vec{l}^2 - (\vec{u}_c(k,p)\cdot\vec{l})^2} \,, 
    \quad \vec{l} = \vec{p} - \vec{k} \,.
    \label{denominator-precise}
\end{align}
It is similar but not equal to the naive denominator factors in \eqref{denominator-naive}. 
What exactly is $\vec{u}_c(k,p)$ in \eqref{denominator-precise}? The vector $\vec{u}_c$ was originally defined for on-shell momenta, 
\begin{align}
    \vec{u}_c = \frac{\vec{P}}{E} = \frac{\vec{p}_1 + \vec{p}_2}{E_1 + E_2} \,.
\end{align}
As we mentioned in \eqref{off-shell-P} and \eqref{off-shell-E}, the total momentum is always conserved but the total energy is not. 
So, we need an off-shell extension for $\vec{u}_c(k,p)$ in $D(k,p)$ 
just as we extended the numerator factor $c_1$. One natural prescription is 
\begin{align}
    \vec{u}_c(k,p) = \vec{P} \left( \frac{E_p + E_k}{2} \right)^{-1} \approx  \left(1 - \frac{E_p - E_k}{2E_p} \right) \vec{u}_c \,,
    \quad 
    \vec{u}_c = \frac{\vec{P}}{E_p} \,.
\end{align} 
(To the leading order in $\Delta E \equiv E_p - E_k$, which is all there is to contribute to the 2PM potential, all prescriptions 
treating $k$ and $p$ on an equal footing give equivalent results.)
Expanding the interpolated denominator with respect to the one fixed at external momenta,  
\begin{align}
\begin{split}
    D(k,p) &\approx \frac{1}{\vec{l}^2 - (1-\Delta E/2E_p)^2 (\vec{u}_c\cdot\vec{l})^2} 
    \\
    &\approx \frac{1}{(\vec{l}^2)_c + (\Delta E/E_p) (\vec{u}_c\cdot\vec{l})^2} 
    \approx \frac{1}{(\vec{l}^2)_c} 
    \left( 1  - \frac{\Delta E}{E_p} \frac{(\vec{u}_c\cdot\vec{l})^2}{(\vec{l}^2)_c} \right) \,.
\end{split}
\end{align}
The $\Delta E$ term leads to the convolution between $\gamma_c/r$ in \eqref{1PM-TL} and a new Fourier integral, 
\begin{align}
     \frac{\gamma_c - \gamma_c^3}{r} = - 4\pi \int_{\vec{q}} \frac{2(\vec{u}_c\cdot\vec{q})^2 }{(\vec{q})^4_c} e^{i\vec{q}\cdot\vec{r}}
    \,, 
     \label{deformed-Fourier-copy}
\end{align}
which contributes to $H^{[2]}_4$ an extra term proportional to $(\gamma_c^2 - \gamma_c^4)$. 
After fixing the constant of proportionality, we 
find that the final result for $H^{[2]}_4$ is
\begin{align}
    H^{[2]}_4 = \gamma_c^2 H^{[2]}_{4,\mathrm{b}} +  2(\gamma_c^2 - \gamma_c^4) H^{[2]}_{4,\mathrm{b}} = (3\gamma_c^2 - 2\gamma_c^4) H^{[2]}_{4,\mathrm{b}} \,.
\end{align}
We refer the readers to appendix~\ref{sec:more-hamiltonian} for further discussions and computations.

\section{Boost} \label{sec:2pm-boost}

We proceed to construct the boost generator compatible with the the Hamiltonian obtained in the previous section. 
Before we begin, we make some technical notes. 

It is useful to introduce a notation to distinguish the two parts of 
$H^{[2]}_4$: 
\begin{align}
    H^{[2]}_4 =  H^{[2]}_{4\alpha} + H^{[2]}_{4\beta} =  (3\gamma_c^2 - 2 \gamma_c^4) H^{[2]}_{4,\mathrm{b}} \,.
\end{align}
The relations between $H^{[2]}_{2,3,4}$ and $(H^{[1]})^2$, which generalize \eqref{1PM-square-bare} to include the dressing factors, will play a crucial role:
\begin{align}
\begin{split}
    H^{[2]}_2 = \frac{E}{m_1 m_2} \frac{4\gamma}{2\gamma^2-1} ( H^{[1]})^2
\,,&\quad 
    H^{[2]}_3 = -\frac{E}{2E_1E_2} (H^{[1]})^2 \,,
    \\
    H^{[2]}_{4\alpha} = \frac{3}{2E} (H^{[1]})^2 \,,&
    \quad 
     H^{[2]}_{4\beta} = -\frac{\gamma_c^2}{E} (H^{[1]})^2 \,.
\end{split}
\label{1PM-square-full}
\end{align}
It is convenient to take components with respect to $\vec{u}_c$ and $\vec{u}_-$ in the $\vec{u}$-space, 
\begin{align}
    \begin{split}
    \vec{u}_c &= z_1 \vec{u}_1 + z_2 \vec{u}_2 \,,
    \\
    \vec{u}_- &= \vec{u}_1 - \vec{u}_2 \,, 
    \end{split}
    \hskip -2cm
    \begin{split}
    \vec{u}_1 &= \vec{u}_c + z_2 \vec{u}_- \,,
    \\ 
    \vec{u}_2 &= \vec{u}_c - z_1 \vec{u}_- \,.
    \end{split}
\end{align}

\paragraph{Poincar\'e algebra up to 1PM}

For a 2-body dynamics without spin, the translation and rotation generators are simply 
\begin{align}
    \vec{P} = \vec{p}_1 + \vec{p}_2 \,,
    \quad 
    \vec{J} =  \vec{x}_1 \times \vec{p}_1 + \vec{x}_2 \times \vec{p}_2 \,, 
    \quad  
    \{ x_i, p_j \} = \delta_{ij} \,.
    \label{P-J}
\end{align}
The complete Poincar\'e algebra reads ($\vec{K} = \vec{G} - t \vec{P}$)
\begin{align}
& \left\{P_i, P_j\right\}= 0 \,, \quad \qquad \left\{P_i, H\right\}=0 \,, \quad \qquad \left\{J_i, H\right\}=0 \,, 
\label{PA-1}
\\
&\left\{J_i, J_j\right\}= \epsilon_{i j k} J_k \,, \quad\left\{J_i, P_j\right\}=\epsilon_{i j k} P_k \,, \quad\left\{J_i, G_j\right\}=\epsilon_{i j k} G_k\,, 
\label{PA-2}
\\
& \left\{G_i, P_j\right\}=\delta_{i j} H \,, \quad\left\{G_i, H\right\}=P_i \,, \quad\qquad \left\{G_i, G_j\right\}=-\epsilon_{i j k} J_k \,.
\label{PA-3}
\end{align}
The non-trivial conditions are all in \eqref{PA-3}, 
which we call ``$H$/$P$/$J$-conditions", respectively. 
In \cite{Lee:2023nkx}, it was shown that at 0PM and 1PM, 
\begin{align}
\begin{split}
    \vec{G}^{[0]} &= H^{[0]} \vec{X}^{[0]} \,,\quad \vec{X}^{[0]} = z_1 \vec{x}_1 + z_2 \vec{x}_2 \,,
    \\
    \vec{G}^{[1]} &= H^{[1]} \vec{X}^{[1]} \,,\quad \vec{X}^{[1]} = z_2 \vec{x}_1 + z_1 \vec{x}_2 \,.
\end{split}
\label{G-0PM-1PM}
\end{align}
and it was conjectured that a similar pattern will persists at higher orders, 
\begin{align}
    \vec{G}^{[n]} =  H^{[n]} \vec{X}^{[n]} \,,
    \quad 
    \vec{X}^{[n]} = \alpha^{[n]}_1 \vec{x}_1 + \alpha^{[n]}_2 \vec{x}_2 \,, 
    \quad 
    \alpha^{[n]}_1 +  \alpha^{[n]}_2 = 1 \,.
    \label{G-X-ansatz}
\end{align}
The restriction on the sum of the two coefficients is sufficient to satisfy the $H$-condition. 
We find in this paper that the ansatz \eqref{G-X-ansatz} is too restrictive and should be modified to allow for more possibilities. 
With hindsight, 
we write our extended ansatz as
\begin{align}
    \vec{G}^{[2]} = H^{[2]}  \vec{X}^{[1]} + Q \vec{r} \,.
     \label{G-X-ansatz-new}
\end{align}
The $H$-condition still holds if $Q$ is translation invariant. 
The appearance of $\vec{X}^{[1]}$ in the ansatz for $\vec{G}^{[2]}$ may look unnatural. 
But, since the difference between any two vectors $\vec{X}$, $\vec{X}'$ of the form in \eqref{G-X-ansatz} 
is proportional to $\vec{r}$, we can choose any such $\vec{X}$ as a reference and absorb the difference 
in the definition of $Q$. To resolve this ambiguity in splitting between $\vec{X}$ and $Q\vec{r}$, 
we demand that $Q$ takes the simplest form possible. 
As we will show shortly, taking $\vec{X}^{[1]}$ as a reference for $\vec{G}^{[2]}$ coincides with the minimal choice for $Q$. 

Before solving the 2PM $P$/$J$-conditions, we recall some of the main features of the 1PM computation. 
In \cite{Lee:2023nkx}, the $P$-condition is reorganized as
\begin{align}
   (\vec{X}^{[0]}-\vec{X}^{[1]} ) \{H^{[0]}, H^{[1]} \} +H^{[0]} \{\vec{X}^{[0]}, H^{[1]} \}+H^{[1]} \{\vec{X}^{[1]}, H^{[0]} \}=0 \,. 
   \label{1PM-P-cond}
\end{align}
The resulting vector can be decomposed into the ``basis" $(\vec{r}, \vec{u}_c ,\vec{u}_-)$. Each component should vanish. 
As for the $J$-condition, the use of the $P$-condition to reorganize the $J$-condition 
simplifies the computation,  
\begin{align}
\begin{split}
   \{ \vec{G}^{[0]} , \vec{G}^{[1]} \}_\times &= H^{[1]} \left[ \{H^{[0]}, \vec{X}^{[1]} \} \times \vec{X}^{[1]} + \{  \vec{G}^{[0]} , \vec{X}^{[1]} \}_\times \right]  \,,
    \\
   \{ \vec{A} , \vec{B} \}_{\times}|_i &\equiv \epsilon_{ijk} \{ A_j , B_k \}  \,.   
\end{split}
\label{1PM-J-cond}
\end{align}

\paragraph{P-condition}

Plugging the ansatz \eqref{G-X-ansatz-new} into the $P$-condition, 
\begin{align}
\vec{\mathcal{P}} \equiv   \{ \vec{G}^{[0]} , H^{[2]} \}  + \{ \vec{G}^{[1]} , H^{[1]} \} +  \{ \vec{G}^{[2]} , H^{[0]} \} = 0 \,,
\end{align}
and rearranging some terms using \eqref{G-0PM-1PM}, we get
\begin{align}
\begin{split}
    \vec{\mathcal{P}} &= \{ H^{[0]}  , H^{[2]} \} (\vec{X}^{[0]} - \vec{X}^{[1]}) + H^{[0]}\{ \vec{X}^{[0]} , H^{[2]} \} + \{ \vec{G}^{[1]} , H^{[1]} \}
    \\
    &\quad + H^{[2]} \{ \vec{X}^{[1]} , H^{[0]} \}  + \{ Q , H^{[0]} \} \vec{r} + Q \{ \vec{r} , H^{[0]} \}
\end{split}
\label{P-cond-all}
\end{align}
To decompose the vector $\vec{\mathcal{P}}$ into components along $\vec{u}_c$, $\vec{u}_-$ and $\hat{n}$, we recall a few facts, 
\begin{align}
\begin{split}
    \vec{X}^{[0]} - \vec{X}^{[1]} = z_{12} \vec{r} \,,
   \quad
   \{ \vec{X}^{[1]} , H^{[0]} \} 
   = \vec{u}_c - z_{12} \vec{u}_- \,,
   \quad 
   \{ \vec{r} , H^{[0]} \} = \vec{u}_- \,.
\end{split}
\end{align}
We also import a few less obvious facts 
from appendix~\ref{sec:more-boost}. 
First, we rewrite the bracket of the 1PM generators in a suggestive form.
\begin{align}
\begin{split}
         \{ \vec{G}^{[1]} , H^{[1]} \} &= A_c \vec{u}_c + A_- \vec{u}_- + A_n \hat{n} \,,
         \\
         A_c &= - \frac{E}{E_1 E_2} (1-3\xi) (H^{[1]})^2 = 2 H^{[2]}_3 + 2H^{[2]}_{4\alpha} \,,
         \\
         A_- &= z_{12} \left[- \left(\frac{E}{m_1m_2}\right)\frac{4\gamma }{2 \gamma^2-1} + \frac{E}{E_1E_2} + \frac{\gamma_c^2 -2}{E} \right]  (H^{[1]})^2 
         \\
         &= z_{12} \left( - H^{[2]}_2  -2 H^{[2]}_3  -\frac{4}{3}  H^{[2]}_{4\alpha} - H^{[2]}_{4\beta}\right) \,,
         \\
         A_n &=  -\bigg(2\gamma_c^2 \left(\hat{n}\cdot \vec{u}_c \right) - z_{12}\Big( \gamma_c^2(\vec{u}_-\cdot\vec{u}_c)_\perp \hat{n}\cdot\vec{u}_c + \hat{n}\cdot \vec{u}_- \Big) \bigg) \frac{(H^{[1]})^2}{E} \\
         &= \frac{z_{12}}{3} \{ H^{[0]} , H^{[2]}_{4\alpha} \} +2 (\hat{n} \cdot \vec{u}_c)H^{[2]}_{4\beta} \,.
\end{split}
\label{1-1-sector-all}
\end{align}
We can write the second term of \eqref{P-cond-all} in a similar way, 
\begin{align}
\begin{split}
 H^{[0]}\{ \vec{X}^{[0]} , H^{[2]} \} &=  B_c \vec{u}_c + B_- \vec{u}_- + B_n \hat{n}\,,
 \\
 B_c &=  -H^{[2]}_1 - H^{[2]}_2- 3H^{[2]}_3- 3H^{[2]}_{4\alpha} - H^{[2]}_{4\beta}\,,
 \\
 B_- &=  z_{12} \left( H^{[2]}_1 + 2 H^{[2]}_2 + 3H^{[2]}_3+2 H^{[2]}_{4\alpha} +2 H^{[2]}_{4\beta} \right) \,. 
 \end{split}
\end{align}
Instead of computing $B_n \hat{n}$ explicitly, we cancel a large part of it against a neighboring term using the identity, 
\begin{align}
    r \{ H^{[0]}  , H^{[2]} \} z_{12}  + B_n  = -  (\vec{u}_c \cdot \vec{\nabla}_r) H^{[2]}   + (\vec{D}_p H^{[2]}  )_n  \,.
    \label{magic-identity}
\end{align}
Here, $(\vec{D}_p F)_n$ defined by the decomposition, 
\begin{align}
 (E_1 \vec{\nabla}_{p_1} + E_2 \vec{\nabla}_{p_2}) F 
 =  (\vec{D}_p F)_c \vec{u}_c +  (\vec{D}_p F)_- \vec{u}_- +  (\vec{D}_p F)_n \hat{n} \,.
\end{align}
Remarkably, for $F \propto \gamma_c^2/r^2$, the RHS of \eqref{magic-identity} vanishes, so  $H^{[2]}_{1,2,3}$ and 
$H^{[2]}_{4\alpha}$ all drop out. 
The contribution from $H^{[2]}_{4\beta}$ is 
\begin{align}
-  (\vec{u}_c \cdot \vec{\nabla}_r) H^{[2]}_{4\beta}   + (\vec{D}_p H^{[2]}_{4\beta}  )_n = -2(\hat{n}\cdot\vec{u}_c) H^{[2]}_{4\beta} \,,
\end{align}
which cancels against a similar term from \eqref{1-1-sector-all}.

In summary, the $P$-condition boils down to 
\begin{align}
\begin{split}
\vec{\mathcal{P}} &=  P_c \vec{u}_c + P_- \vec{u}_- + P_n \hat{n}\,,
 \\
 P_c &= 0 \,,
 \quad 
 P_- = -\frac{1}{3} z_{12} H^{[2]}_{4\alpha} + Q \,,
 \quad 
 P_n = \frac{z_{12}}{3} \{ H^{[0]} , H^{[2]}_{4\alpha} \}- \{ H^{[0]} , Q \} \,.
 \end{split}
\end{align}
Clearly, the unique solution to the $P$-condition is 
\begin{align}
    Q = \frac{z_{12}}{3} H^{[2]}_{4\alpha} = z_{12} \gamma_c^2 H_{4,\mathrm{b}} = z_{12} \frac{(H^{[1]})^2}{2E} \,.
    \label{Q-search-done}
\end{align}

\paragraph{J-condition}

While we solve the $J$-condition, 
\begin{align}
    \vec{\mathcal{J}} &\equiv  \{ \vec{G}^{[2]} , \vec{G}^{[0]} \}_{\times} + \frac{1}{2} \{ \vec{G}^{[1]} , \vec{G}^{[1]} \}_{\times}  = 0 \,,
\end{align}
we keep track of the two terms in our main ansatz ($\vec{G}^{[2]} = H^{[2]} \vec{X}^{[1]} + Q \vec{r}$) separately. 
For the first term, we note that 
\begin{align}
\begin{split}
      \{ \vec{G}^{[0]},  H^{[2]} \vec{X}^{[1]}  \}_\times 
      &= \{ \vec{G}^{[0]},  H^{[2]}  \} \times  \vec{X}^{[1]} +   H^{[2]}  \{ \vec{G}^{[0]}, \vec{X}^{[1]}  \}_\times 
      \\
      &=  - \left[ \{ \vec{G}^{[1]},  H^{[1]}  \} + \{ \vec{G}^{[2]},  H^{[0]}  \}  \right]\times  \vec{X}^{[1]} 
    -  H^{[2]}  \{ H^{[0]}, \vec{X}^{[1]}  \} \times \vec{X}^{[1]}  
    \\
    &= - H^{[1]}\{ \vec{X}^{[1]} , H^{[1]}\}\times \vec{X}^{[1]} 
    - \{  Q \vec{r} , H^{[0]} \} \times \vec{X}^{[1]} 
    \,.
\end{split}
\end{align}
To reach the second line, we used the 2PM $P$-condition as well as the \emph{1PM $J$-condition} \eqref{1PM-J-cond}. In the last step, we used  \eqref{G-X-ansatz-new} once again and also used $\vec{X}^{[1]}\times \vec{X}^{[1]} = 0$.

The sum of all $Q$-independent terms gives
\begin{align}
\begin{split}
      - H^{[1]}\{ \vec{X}^{[1]} , H^{[1]}\}\times \vec{X}^{[1]}  + \frac{1}{2} \{ \vec{G}^{[1]}, \vec{G}^{[1]}\}_\times
      = \frac{1}{2} (H^{[1]})^2  \{ \vec{X}^{[1]}, \vec{X}^{[1]}\}_\times \,.
\end{split}
\end{align}
The sum of all $Q$-dependent terms gives
\begin{align}
\{ \vec{G}^{[0]},  Q \vec{r} \}_\times- \{  Q \vec{r} , H^{[0]} \} \times \vec{X}^{[1]} = z_{12}  Q \{  \vec{r} , H^{[0]} \} \times \vec{r} + H^{[0]} \{ Q\vec{r} , \vec{X}^{[0]} \}_\times \,.
\end{align}
The full $J$-condition is then
\begin{align}
\begin{split}
    \vec{\mathcal{J}} &= \frac{1}{2} (H^{[1]})^2  \{ \vec{X}^{[1]}, \vec{X}^{[1]}\}_\times + z_{12}  Q (\vec{u}_- \times \vec{r}) + H^{[0]} \{ Q\vec{r} , \vec{X}^{[0]} \}_\times
    \\
    &\equiv \vec{\mathcal{J}}_A + \vec{\mathcal{J}}_B + \vec{\mathcal{J}}_C \,.
\end{split}
\end{align}
Using the fact that 
\begin{align}
     \{ \vec{X}^{[1]} , \vec{X}^{[1]} \}_{\times} &= -\frac{2}{E}  (z_2^2 \vec{u}_1 - z_1^2 \vec{u}_2) \times \vec{r} 
     = \frac{2}{E}(z_{12} \vec{u}_c - (1-3\xi) \vec{u}_- ) \times \vec{r}  \,, 
\end{align}
and the last equality in \eqref{Q-search-done}, we get
\begin{align}
    \vec{\mathcal{J}}_A  = 2Q \left(\vec{u}_c - \frac{1}{z_{12}} (1-3\xi) \vec{u}_- \right) \times \vec{r} \,.
\end{align}
It remains to compute 
\begin{align}
    \vec{\mathcal{J}}_C  = H^{[0]} \{ \vec{X}^{[0]} , Q\vec{r} \}_\times
    = H^{[0]} \{ \vec{X}^{[0]} , Q\} \times \vec{r} + Q H^{[0]} \{ \vec{X}^{[0]} , \vec{r} \}_\times \,.
\end{align}
Further computations show  
\begin{align}
\begin{split}
      H^{[0]} \{ \vec{X}^{[0]} , \vec{r} \}_\times &= (\vec{u}_c -z_{12} \vec{u}_-) \times \vec{r} \,,
\\
    H^{[0]} \{ \vec{X}^{[0]} , Q\}_u &= \left(-3\vec{u}_c +   \frac{2( 1- \xi)}{z_{12}} \right) Q \,.  
\end{split}
\end{align}
Combining everything, and checking the coefficients of $(\vec{u}_c \times \vec{r})$ and $(\vec{u}_- \times \vec{r})$ separately, we confirm that the $J$-condition, $\vec{\mathcal{J}} = 0$, holds.

\section{Discussion} \label{sec:discussion}

In the spinless point particle limit, we succeeded in constructing the 2PM Hamiltonian in an arbitrary lab frame and the 2PM boost generator, thereby explicitly verifying the global Poincar\'e algebra up to the 2PM order. We did not attempt to compare our result with the overlapping result in the PN expansion in \cite{Damour:2000kk} and many subsequent work. Since the PM computation is exact in velocities, the expressions for the PM generators tend to be  more compact than their PN counterparts. 

One obvious extension of this work would be to proceed to the 3PM order. 
At 2PM, we had to separately keep track of the four terms comprising the Hamiltonian. For all but one terms, 
the interplay between the 2PM Hamiltonian and the iteration of the 1PM Hamiltonian played a vital role. 
A cursory look at the 3PM COM Hamiltonian \cite{Bern:2019nnu,Bern:2019crd} shows that the number of terms is at least doubled and that the iteration includes many terms such as $c_1^3$, $c_1(c_1')^2$, $c_1^2 c_1''$, $c_2c_1$, $c_2c_1'$ and so on. Here, $c_1$ and $c_2$ are the numerator factors of $H^{[1]}$ and $H^{[2]}_1$, respectively. 
The appearance of a second order derivative as in $c_1''$ would require a more precise prescription for the interpolation during the off-shell extension. At 2PM, we needed at most $c_1'$ and 
the difference between different prescriptions was immaterial. 

Another important extension would be to study $N$-body dynamics. 
Take a 3-body Hamiltonian for an example. 
As explained in \cite{Jones:2022aji}, while the 1PM Hamiltonian is merely the sum of pairwise interactions, the 2PM Hamiltonian includes a genuine 3-body interaction term. How this term affects the global Poincar\'e algebra is an interesting question. 

Even for a binary system at 2PM, including the spin effects is a challenging open problem. The 2PM amplitudes to all orders in spin were recently proposed in \cite{Alessio:2023kgf,Aoude:2023vdk}. But, extra steps are needed to map the amplitudes to the Hamiltonian. In particular, the Thomas-Wigner rotation factor among an incoming momentum, an outgoing momentum and the common reference frame for the binary should be included. At 1PM, the complete form of the rotation factor was first given in the COM frame in \cite{Chung:2020rrz} and was generalized to lab frames in \cite{Lee:2023nkx}. 
It would be interesting to see how the attempt to construct the boost generator may constrain the form of the rotation factor at 2PM.

\vskip 1cm 

\acknowledgments

The work of HL and SL is supported by the National Research Foundation of Korea grant NRF-2019R1A2C2084608. 
The work of KL is supported by an appointment to the JRG
Program at the APCTP through the Science and Technology Promotion Fund and Lottery Fund of the Korean Government. 
KL is also supported by the National Research Foundation
of Korea grant funded by the Korean government(MSIT) No.RS-2023-00249451 and the Korean Local Governments of Gyeongsangbuk-do Province and Pohang City.
We are grateful to Sungjay Lee for discussions. 
 HL and SL thank the Asia Pacific Center for Theoretical Physics for hospitality where a part of this work was done. 
 HL and KL thank Korea Institute for Advanced Study for hospitality where a part of this work was done.

\newpage 
\appendix

\section{More on the Hamiltonian} \label{sec:more-hamiltonian}

In this appendix, we review the Born subtraction contributing to the 2PM Hamiltonian in the COM frame, 
and see what new features arise as we move to a lab frame.

\paragraph{COM frame}

We follow \cite{Cheung:2018wkq,Cristofoli:2019neg}.
An intermediate step involves the integral,
\begin{align}
\mathcal{I}_\mathrm{B} = 
    \int_{\vec{k}} \frac{c_1^2\left(\frac{p^2+k^2}{2}\right)}{\left(E_p-E_k+i \epsilon\right)(\vec{p}-\vec{k})^2 (\vec{p}'-\vec{k})^2} \,.
    \label{Born-COM}
\end{align}
A key idea in the next step is that the super-classical and classical terms all come from near the pole at $E_p - E_k = 0$. 
For the time being, let us focus on the factor 
\begin{align}
    \mathcal{F} = \frac{c_1^2\left(\frac{p^2+k^2}{2}\right)}{E_p-E_k+i \epsilon} \,, 
    \qquad 
    c_1(p^2) \propto \frac{2\gamma^2-1}{E_1 E_2} \,. 
\end{align}
The references suggest that we use $k^2$ as an independent variable and expand both the denominator and the numerator,
\begin{align}
    E_p - E_k &= \left. \frac{d E}{d (k^2)} \right|_{k^2 = p^2} (p^2 -k^2) - \frac{1}{2} \left. \frac{d^2 E}{d (k^2)^2}\right|_{k^2 = p^2} (p^2 -k^2)^2 + \cdots \,,
    \\
    c_1^2\left(\frac{p^2+k^2}{2}\right) &= c_1^2(p^2) + (k^2-p^2) c_1(p^2) \left. \frac{d c_1(k^2)}{dk^2} \right|_{k^2 = p^2} + \cdots \,.
\end{align}
The super-classical term arises from the leading $\mathcal{O}(\hbar^{-1})$ term, 
\begin{align}
\mathcal{F}_{-1} =  \left( \frac{1}{E'} \right) \frac{c_1(p^2)^2}{p^2 - k^2 + i\epsilon } \,, 
\quad 
    f' \equiv  \left. \frac{df}{d(k^2)} \right|_{k^2 =p^2} \,. 
    \label{COM-super}
\end{align}
More important to us is the classical term, which comes from the next-to-leading terms. 
\begin{align}
   \mathcal{F}_0 = - \frac{c_1 c_1'}{E'}  + \frac{E''}{2(E')^2} (c_1)^2  \,.
   \label{born-classical}
\end{align}
Using the facts, 
\begin{align}
\begin{split}
&E' = \frac{1}{2} \left( \frac{1}{E_1} + \frac{1}{E_2} \right) = \frac{1}{2E\xi} \,,
    \quad 
    E'' = -\frac{E^3}{4E_1^3 E_2^3} (1 - 3\xi) \,,
    \\
    &c_1' = \frac{4 \gamma \gamma'}{E_1 E_2}  - c_1 \frac{E^2}{2E_1^2 E_2^2} (1-2\xi) \,, \quad 
    \gamma' = \frac{1}{2 m_1m_2 \xi} \,, 
\end{split}
\label{born-com-facts}
\end{align}
we obtain 
\begin{align}
\begin{split}
    - \frac{c_1 c_1'}{E'}  &= \frac{E}{E_1E_2} \left( -4 \frac{\gamma c_1 }{m_1 m_2}   +  (1-2\xi) (c_1)^2 \right) \,,
    \\ 
    \frac{E''}{2(E')^2} (c_1)^2 &= - \frac{E}{2E_1E_2} (1-3\xi)(c_1)^2 \,,
\end{split}
\end{align}
and finally,  
\begin{align}
    \mathcal{F}_0 = \frac{E}{2E_1E_2} \left[ -8 \frac{\gamma c_1}{m_1m_2}  + (1-\xi)(c_1)^2  \right] \,.
\end{align}
The $\gamma c_1$, $(c_1)^2$, $\xi(c_1)^2$ terms correspond to $H^{[2]}_{2,\mathrm{b}}$, $H^{[2]}_{3,\mathrm{b}}$, $H^{[2]}_{4,\mathrm{b}}$ in \eqref{H-2PM-COM}, respectively. 

\paragraph{Lab frame}

In the COM frame, $\vec{p}_1 = \vec{p} = - \vec{p}_2$ implies that $E_1$ and $E_2$ depend only on $\vec{p}^2$. The level surface of $E = E_1 +E_2 $ is a sphere. In the lab frame, the situation is more complicated. 

Let us first examine the Born denominator $(E_p-E_k)^{-1}$. 
As we vary $\vec{p}_1$ and $\vec{p}_2$ while keeping $\vec{P} = \vec{p}_1 + \vec{p}_2$ fixed, 
the condition for the total energy staying constant is 
\begin{align}
    \sqrt{\vec{p}_1^2 +m_1^2} + \sqrt{\vec{p}_2^2 +m_2^2} 
    =  \sqrt{(\vec{p}_1-\vec{l})^2 +m_1^2} + \sqrt{(\vec{p}_2+\vec{l})^2 +m_2^2} \,.
\end{align}
Squaring both sides and simplifying a bit, we find an equation for an ellipsoid, 
\begin{align}
    f_{\vec{p}}(\vec{l}) \equiv  \vec{l}^2 - (\vec{u}_c\cdot \vec{l})^2 -\frac{2E_1 E_2}{E} ( \vec{u}_- \cdot \vec{l} )= 0 \,.
    \label{l-ellipsoid}
\end{align}
This expression will be useful in extracting the super-classical and classical terms from the Born subtraction in the lab frame. 
Expanding the Born denominator in $\vec{l} = \vec{p} - \vec{k}$,  we find 
\begin{align}
\begin{split}
       E_p - E_k 
       &= -\frac{1}{2E\xi} f(\vec{l}) \left( 1 - \frac{z_{12}}{E \xi} (\vec{u}_c\cdot\vec{l}) \right) + \left(\frac{1}{2E\xi } \right)^3 (1 -3 \xi ) f(\vec{l})^2  + \mathcal{O}(l^3) \,.
\end{split}
\label{Ep-Ek}
\end{align}

The numerator $c_1$ consists of two distinct factors, $(2\gamma^2-1)$ and $1/(E_1E_2)$. For the first factor, we note that the following relation holds in any frame:
\begin{align}
    E^2 - \vec{P}^2 = m_1^2 + m_2^2 + 2m_1 m_2 \gamma \,.
\end{align}
If we vary $E$ while keeping $\vec{P}$ fixed, we find 
\begin{align}
    E dE  = m_1 m_2 d\gamma 
    \quad 
    \Longrightarrow
    \quad 
    \frac{d\gamma}{dE} = \frac{E}{m_1 m_2} \,.
    \label{dgamma-dE}
\end{align}
When restricted to the COM frame, it agrees with $\gamma'/E'$ in \eqref{born-com-facts}, but we stress that \eqref{dgamma-dE} is valid in any frame. 
This simple argument proves that $H^{[2]}_2$ receives nothing more than the simple dressing by $\gamma_c^2$.

The $1/(E_1E_2)$ factor in $c_1$ deserves more attention.
To distinguish the variance through the change in the total energy $E$ from the variance independent of the change in $E$, 
we propose a double-expansion in $f(\vec{l})$ and $(\vec{u}_c\cdot \vec{l})$:
\begin{align}
\begin{split}
    \frac{E_1E_2}{E_1(\vec{p}_1 - \vec{l})E_2(\vec{p}_2 + \vec{l})} 
    = 1 &- \frac{z_{12}}{E\xi}  ( \vec{u}_c\cdot\vec{l}) +  \frac{(1-3\xi)}{(E\xi)^2}   ( \vec{u}_c\cdot\vec{l})^2 
    \\
    & - \frac{1}{2(E\xi)^2}f(\vec{l})\left[ (1-2\xi)  - (3-4\xi) \frac{z_{12}}{E\xi}  (\vec{u}_c \cdot \vec{l})  \right] 
    \\
    &  + \frac{1}{8(E\xi)^4}f(\vec{l})^2 (3-12\xi+8\xi^2) + \mathcal{O}(l^3)\,. 
\end{split}
\label{E1E2-expand}
\end{align}
At each order in $f(\vec{l})$, the leading coefficient independent of $(\vec{u}_c\cdot \vec{l})$ should agree with that computed in the COM frame. 

In \eqref{Ep-Ek} and \eqref{E1E2-expand}, we observed factors depending on $(\vec{u}_c\cdot \vec{l})$, possibly 
signaling a new feature as we move away from the COM frame.  
However, since $\vec{l}$ scales as $\mathcal{O}(\hbar)$ in the classical limit, most of them become irrelevant. 
The only possible exception is the one contributing to the super-classical term \eqref{COM-super}. 
Fortunately, the leading $(\vec{u}_c\cdot \vec{l})$ corrections to \eqref{Ep-Ek} and \eqref{E1E2-expand} cancel out precisely, 
so even the super-classical term remains unaffected.

\paragraph{Box diagram} 

According to \cite{Bjerrum-Bohr:2021vuf}, in $D=4-2\epsilon$ dimensions, the sum of the box and the crossed-box diagrams gives (neglecting quantum corrections) 
\begin{align}
\begin{split}
       \mathcal{I}_{\boxtimes} &= \frac{1}{|q|^{2+2 \epsilon}} \frac{(4\pi)^{\epsilon} }{32m_1 m_2} 
    \left(
    -\frac{\Gamma(1+\epsilon) \Gamma(-\epsilon)^2}{ \pi \Gamma(-2 \epsilon) \sqrt{\gamma^2-1}} 
    \right. 
    \\
    &\hskip 4cm 
    \left. 
    +\frac{i\left(m_1+m_2\right)|q|}{m_1 m_2\left(\gamma^2-1\right)} \frac{\Gamma\left(\frac{1}{2}-\epsilon\right)^2 \Gamma\left(\frac{1}{2}+\epsilon\right)}{\pi^{3/2} \Gamma(-2 \epsilon)}\right) \,.
\end{split}
\label{1-loop-box-final}
\end{align}
The first term is the infrared-divergent super-classical term and the second term is the finite classical term. 
A key feature of \eqref{1-loop-box-final} is Lorentz invariance. In the lab frame, the $|q|$ in \eqref{1-loop-box-final} should be understood as 
$|q|^2 = (\vec{q})^2_c = \vec{q}^2 - (\vec{u}_c\cdot \vec{q})^2$. 
Another important fact is that in four space-time dimensions $(\epsilon = 0)$, the classical term vanishes:
\begin{align}
    \frac{\Gamma\left(\frac{1}{2}-\epsilon\right)^2 \Gamma\left(\frac{1}{2}+\epsilon\right)}{\pi^{\frac{3}{2}} \Gamma(-2 \epsilon)} 
    = 0 - 2\epsilon + \mathcal{O}(\epsilon^2) \,.
\end{align}
In summary, \eqref{1-loop-box-final} is equally valid in any reference frame. 
Nevertheless, we find it instructive to compute the super-classical term in a seemingly non-relativistic way. 

The relevant loop integral is the scalar box integral, 
\begin{align}
\begin{split}
     \mathcal{I}_{\square} &= \int \frac{d^D \ell}{(2\pi)^D} \frac{1}{((p_1 -\ell)^2 + m_1^2 -i\varepsilon)((p_2 +\ell)^2 +m_2^2 -i\varepsilon) \ell^2 (q -\ell)^2} 
     \\
     &= \int \frac{d^D \ell}{(2\pi)^D} \frac{1}{(\ell^2 -2p_1\cdot \ell  -i\varepsilon)(\ell^2 + 2p_2 \cdot \ell  -i\varepsilon) \ell^2 (q -\ell)^2} \,.
\end{split}
\end{align}
We work in the $(-+++)$ metric signature. Evaluating this integral, even approximately, is a great challenge, but it is relatively easy to isolate the leading super-classical term. 
One simply replaces the two massive propagators by delta functions in view of the relation 
\begin{align}
    \frac{1}{x-i\varepsilon} = \mathrm{p.v.}\left(\frac{1}{x}\right) + \pi i \delta(x) \,.
\end{align}
Then the integral becomes
\begin{align}
\begin{split}
     \mathcal{I}_{\square} &\approx \pi i \int \frac{d^D \ell}{(2\pi)^D} \frac{\delta (\ell^2 -2p_1\cdot \ell) \delta(\ell^2 + 2p_2 \cdot \ell)}{ \ell^2 (q -\ell)^2} 
     \\
     &= \frac{\pi i}{2} \int \frac{d^D \ell}{(2\pi)^D} \frac{\delta ((p_1+p_2)\cdot \ell) \delta(\ell^2 -(p_1-p_2) \cdot \ell)}{ \ell^2 (q -\ell)^2}
     \\
     &= \frac{i}{4E} \int \left. \frac{d^d \vec{l}}{(2\pi)^d} \frac{\delta(\ell^2 -(p_1-p_2) \cdot \ell)}{ \ell^2 (q -\ell)^2} \right|_{\ell^0 = \vec{u}_c\cdot \vec{l}} \,.
\end{split}
\end{align}
One of the delta functions led to the replacement $\ell^0 = \vec{u}_c\cdot \vec{l}$, which in turn implies 
\begin{align}
    \ell^2 \rightarrow (\vec{l}^2)_c \,,
    \quad 
    (q- \ell)^2 \rightarrow (\vec{q}-\vec{l})^2_c \,,
    \quad 
    \ell^2 -(p_1-p_2) \cdot \ell \rightarrow f_{\vec{p}}(\vec{l}) \,.
\end{align}
Hence the super-classical term from the box integral takes the same form as the corresponding term 
in the Born integral \eqref{Born-COM} not only in the COM frame but in any lab frame.

\paragraph{Fourier transform}

We use the shorthand notations, 
\begin{align}
    \int_{\vec{r}} \equiv \int d^3\vec{r} \,,
    \qquad 
    \int_{\vec{q}} \equiv \int \frac{d^3\vec{q}}{(2\pi)^3}  \,.
\end{align}

Before deformation from the sphere to the ellipsoid in the 
$q$-space, the Fourier transform of the $(1/r)$ potential is well-known:
\begin{align}
    \frac{1}{r} = 4\pi  \int_{\vec{q}} \frac{e^{i\vec{q}\cdot\vec{r}}}{q^2}  
    \quad 
    \Longleftrightarrow
    \quad 
    \frac{4\pi}{q^2} =   \int_{\vec{r}} \frac{e^{-i\vec{q}\cdot\vec{r}}}{r}  \,.
\end{align}
The transform of the $(1/r^2)$ potential comes for free. We simply switch the labels between $\vec{r}$ and $\vec{q}$ and adjust 
the factors of $(2\pi)$. The result is 
\begin{align}
    \frac{1}{r^2} = 2\pi^2  \int_{\vec{q}} \frac{e^{i\vec{q}\cdot\vec{r}}}{q}  
    \quad 
    \Longleftrightarrow
    \quad 
    \frac{2\pi^2}{q} =   \int_{\vec{r}} \frac{e^{-i\vec{q}\cdot\vec{r}}}{r^2}  \,. 
\end{align}
Alternatively, we can approach the $(1/r^2)$ potential via a convolution, 
\begin{align}
     \frac{1}{r^2} = (4\pi)^2 \int_{\vec{q}} \int_{\vec{l}} \frac{e^{i\vec{q}\cdot\vec{r}}}{\vec{l}^2 (\vec{q}-\vec{l})^2} \,.
\end{align}
The two approaches agree through the relation
\begin{align}
    \int_{\vec{l}} \frac{1}{\vec{l}^2 (\vec{q}-\vec{l})^2} = \frac{1}{8q} \,.
\end{align}

The ellipsoid deformation of the lab frame induces an  an effective metric in the $q$-space:
\begin{align}
    (g_c)_{ij} = \delta_{ij} - (u_c)_i (u_c)_j 
    \quad 
    \Longleftrightarrow
    \quad 
    (g_c)^{ij} = \delta^{ij} + \frac{(u_c)^i (u_c)^j}{1-\vec{u}_c^2} \,.
\end{align}
The deformed Fourier transforms can be performed with the help of orthonormal frames associated with the metric. 
The results for the $(1/r)$ and $(1/r^2)$ potentials are  
\begin{align}
    \frac{\gamma_c}{r} = 4\pi  \int_{\vec{q}} \frac{e^{i\vec{q}\cdot\vec{r}}}{(\vec{q}^2)_c} \,,
    \qquad 
    \frac{\gamma_c^2}{r^2} = 2\pi^2  \gamma_o \int_{\vec{q}} \frac{e^{i\vec{q}\cdot\vec{r}}}{(\vec{q}^2)_c^{1/2}} \,.
    \label{deformed-Fourier-1}
\end{align}
The convolution argument is deformed accordingly, 
\begin{align}
     \int_{\vec{l}} \frac{1}{(\vec{l}^2)_c (\vec{q}-\vec{l})^2_c } = \frac{\gamma_o}{8 (\vec{q}^2)_c^{1/2} } \,.
     \label{deformed-covolution-1}
\end{align}
Finally, to compute the extra contribution to $H^{[2]}_4$, we also need 
\begin{align}
     \frac{\gamma_c - \gamma_c^3}{r} = - 4\pi \int_{\vec{q}} \frac{2(\vec{u}_c\cdot\vec{q})^2 }{(\vec{q}^2)_c^4} e^{i\vec{q}\cdot\vec{r}}
    \,.
     \label{deformed-Fourier-4}
\end{align}
Again, one can verify it using the orthonormal frames and direct integration.

\newpage 
\section{More on the boost generator} \label{sec:more-boost}

Recall that the $P$-condition is 
\begin{align}
    \{ \vec{G}^{[2]} , H^{[0]} \} + \{ \vec{G}^{[1]} , H^{[1]} \} + \{ \vec{G}^{[0]} , H^{[2]} \} = 0 \,.
\end{align}
Each term will produce terms proportional to $\vec{u}_{1,2}$ or  $\vec{x}_{1,2}$. 

\paragraph{(1-1) part}

This part does not rely on the 2PM ansatz: 
\begin{align}
\begin{split}
     \{ \vec{G}^{[1]} , H^{[1]} \} &= \{  H^{[1]} \vec{X}^{[1]} , H^{[1]} \}
     =  H^{[1]}  \{ \vec{X}^{[1]} , H^{[1]} \} = \frac{1}{2}   \{ \vec{X}^{[1]} , (H^{[1]})^2 \} \,.
\end{split}
\end{align}
It is useful to separate the $\gamma_c$ factor, 
\begin{align}
    H^{[1]} = \gamma_c H^{[1]}_{\mathrm{b}}  \,.
\end{align}
The bare part is easy to compute:
\begin{align}
\begin{split}
       \{ \vec{X}^{[1]} , H^{[1]}_{\mathrm{b}} \} &= 
    \left[ -\frac{E}{E_1E_2} (z_2^2\vec{u}_1 + z_1^2 \vec{u}_2) - \frac{4\gamma}{2\gamma^2-1} \frac{Ez_{12}}{m_1m_2} \vec{u}_- 
    - \frac{\hat{n}}{E}  (\vec{w}\cdot\hat{n}) \right] H^{[1]}_{\mathrm{b}} \,,
    \\
    \vec{w} &\equiv z_2 \vec{u}_1 + z_1 \vec{u}_2 = \{ \vec{X}^{[1]} , H^{[0]} \} = E \{ \vec{r} , z_1 \} \,.
\end{split}
\end{align}
The $\gamma_c$ factor requires more work. The result is 
\begin{align}
\begin{split}
  \{\vec{X}^{[1]}, \gamma_{c}\} 
   & = -\frac{\gamma _c^3}{E} \Big(\hat{n} \big(\hat{n}\cdot \vec{u}_c\big) \big(1 + \left( \vec{u}_c \cdot \vec{w} \right)_{\bot} \big)+\left(\vec{u}_c\right)^2_{\bot}\vec{w} -\vec{u}_c\Big) \,, 
   \\
   (\vec{a}\cdot\vec{b})_\bot &\equiv \vec{a}\cdot\vec{b} - (\vec{a}\cdot\hat{n})(\hat{n}\cdot \vec{b}) \,. 
\end{split}
\end{align}
Combining everything and simplifying a bit, we find 
\begin{align}
\begin{split}
    &\{ \vec{X}^{[1]} , H^{[1]} \} 
    \\
    &= z_{12} \vec{u}_- \left(\frac{\gamma_c^2 }{E}- \left(\frac{E}{m_1m_2}\right)\frac{4\gamma }{2 \gamma^2-1} \right) H^{[1]} 
   -\left(\frac{z_2^2}{E_1}\vec{u}_1+\frac{z_1^2}{E_2}\vec{u}_2\right)H^{[1]}
  \\
  &\qquad - \hat{n} \bigg(\gamma_c^2 \left(\hat{n}\cdot \vec{u}_c \right) \big(1 + \left( \vec{u}_c \cdot\vec{w} \right)_{\bot} \big) + \hat{n}\cdot \vec{w} \bigg) \frac{H^{[1]}}{E} \,.
\end{split}
\end{align}
Switching to the $\vec{u}_c$, $\vec{u}_-$ basis, we obtain 
\begin{align}
    \begin{split}
        &\{ \vec{X}^{[1]} , H^{[1]} \} 
        \\
        &= z_{12} \vec{u}_- \left[\frac{\gamma_c^2 }{E}- \left(\frac{E}{m_1m_2}\right)\frac{4\gamma }{2 \gamma^2-1} + \frac{E}{E_1E_2}(1-2\xi) \right] H^{[1]} 
        - \vec{u}_c \frac{E}{E_1E_2}(1-3\xi) H^{[1]}
      \\
      &\qquad - \hat{n}  \bigg(2\gamma_c^2 \left(\hat{n}\cdot \vec{u}_c \right) - z_{12}\Big( \gamma_c^2(\vec{u}_-\cdot\vec{u}_c)_\perp \hat{n}\cdot\vec{u}_c + \hat{n}\cdot \vec{u}_- \Big) \bigg) \frac{H^{[1]}}{E} \,.
    \end{split}
    \end{align}

\paragraph{(0-2) part} 

When we compute $H^{[0]} \{\vec{X}^{[0]},H^{[2]}\}$, 
the terms proportional to $\vec{u}_{1,2}$ come from 
\begin{align}
H^{[0]} \{\vec{X}^{[0]},H^{[2]}\}_u = \left. \vec{D}_p H^{[2]} \right|_u \equiv   \left. ( E_1 \vec{\nabla}_{p_1} + E_2 \vec{\nabla}_{p_2} ) H^{[2]} \right|_u \,.
\end{align}
The four pieces of $H^{[2]}$ depend on $\vec{p}_{1,2}$ either through 
powers of $\gamma_c$ and $\gamma_o$ or through $F_{1,2,3,4}$. 
The $\gamma$ factor does not contribute since $\{ \vec{G}^{[0]} , \gamma \} = 0$ 
as discussed in \cite{Lee:2023nkx}. 
Applying chain rules and projecting onto $\vec{u}_c$ and $\vec{u}_-$, 
we find an appealing intermediate expression: 
\begin{align}
\begin{split}
   H^{[0]} \{\vec{X}^{[0]},H^{[2]}\}_u &=
   \vec{u}_c \left(\gamma_c \frac{\partial}{\partial \gamma_c}  +\gamma_o \frac{\partial}{\partial \gamma_o } + E_1 \frac{\partial}{\partial E_1}  + E_2 \frac{\partial}{\partial E_2} \right) H^{[2]}
   \\
   &\qquad \qquad + \vec{u}_-  \left( \frac{E_1 E_2}{E} \right) \left( \frac{\partial}{\partial E_1} - \frac{\partial}{\partial E_2}\right) H^{[2]} \,.
\end{split}
\end{align}
To compute the $\vec{x}$ terms from $H^{[0]} \{\vec{X}^{[0]},H^{[2]}\}$, we derive an identity that holds 
for any function of $\vec{p}_{1,2}$ and $\vec{r}$: 
\begin{align}
\begin{split}
        H^{[0]} \{ \vec{X}^{[0]} , F \}_x &= E \{ z_1 , F \} \vec{x}_1 +  E \{ z_2 , F \} \vec{x}_1 + (\vec{D}_p F)_x
        \\
        &= E \{ z_1 , F \} \vec{r} + (\vec{D}_p F)_x
        \\
        &= - \left[ E (\vec{\nabla}_{p_1} z_1 - \vec{\nabla}_{p_2} z_1) \cdot \vec{\nabla}_r F \right] \vec{r} + (\vec{D}_p F)_x
        \\
        &= - \left[ (z_2 \vec{u}_1 + z_1 \vec{u}_2) \cdot \vec{\nabla}_r F \right]\vec{r} + (\vec{D}_p F)_x
        \\
        &= - \left[ (\vec{u}_c \cdot \vec{\nabla}_r) F - z_{12} (\vec{u}_- \cdot \vec{\nabla}_r) F \right] \vec{r} + (\vec{D}_p F)_x
        \\
        &= - \left[ (\vec{u}_c \cdot \vec{\nabla}_r) F \right] \vec{r}  - z_{12} \{H^{[0]} , F \} \vec{r} + (\vec{D}_p F)_x \,.
\end{split}
\end{align}
The Poisson bracket with $H^{[0]}$ acts as a derivative in the $\vec{r}$-space:
\begin{align}
    \{ H^{[0]}, Z \} = - \vec{u}_- \cdot \vec{\nabla}_r Z \equiv - D_- Z  \,.
\end{align}
As such, it does not affect functions of $\vec{p}_{1,2}$ only. 
\begin{align}
     \{ H^{[0]}, \gamma_c^2 \} = \frac{2\gamma_c^4}{r} (\hat{n} \cdot \vec{u}_c) (\vec{u}_- \cdot \vec{u}_c)_{\perp} \,,
     \quad 
     \{ H^{[0]}, r^{-2} \} = \frac{2\hat{n}\cdot \vec{u}_-}{r^3} \,.
\end{align}
Since $H^{[2]}_{1,2,3,4\alpha}$ are all proportional to $\gamma_c^2/r^2$, we have 
\begin{align}
     \{ H^{[0]}, H^{[2]}_i\} = \frac{2 H^{[2]}_i}{r} \Big(\gamma _c^2 \left(\vec{u}_-\cdot \vec{u}_c\right)_{\bot} (\hat{n} \cdot \vec{u}_c) +\hat{n}\cdot \vec{u}_-\Big) 
     \qquad (i=1,2,3,4\alpha) \,.
\end{align}
As for $H^{[2]}_{4\beta}$, we find 
\begin{align}
     \{ H^{[0]}, H^{[2]}_{4\beta} \} = \frac{2 H^{[2]}_{4\beta}}{r} \Big(2\gamma _c^2 \left(\vec{u}_-\cdot \vec{u}_c\right)_{\bot} (\hat{n} \cdot \vec{u}_c) +\hat{n}\cdot \vec{u}_-\Big) \,.
\end{align}

\newpage 
\bibliographystyle{JHEP}
\bibliography{reference}

\end{document}